\documentclass[conference]{IEEEtran}
\IEEEoverridecommandlockouts

\usepackage{cite}
\usepackage{amsmath,amssymb,amsfonts}
\usepackage{algorithmic}
\usepackage{graphicx}
\usepackage{textcomp}
\usepackage{xcolor}
\usepackage{hyperref}
\usepackage{multirow}
\usepackage{subcaption}
\usepackage{array}
\def\BibTeX{{\rm B\kern-.05em{\sc i\kern-.025em b}\kern-.08em
    T\kern-.1667em\lower.7ex\hbox{E}\kern-.125emX}}
\begin{document}

\title{Differential Transformer-driven 6G Physical Layer for Collaborative Perception Enhancement}

\author{\IEEEauthorblockN{Soheyb Ribouh}
\IEEEauthorblockA{\textit{Univ Rouen Normandie}\\ 
\textit{LITIS UR 4108} \\
\textit{ F-76000 Rouen, France} \\
}
\and
\IEEEauthorblockN{Osama Saleem}
\IEEEauthorblockA{\textit{INSA Rouen Normandie}\\
\textit{LITIS UR 4108} \\
\textit{ F-76000 Rouen, France} \\
}
\and
\IEEEauthorblockN{Mohamed Ababsa}
\IEEEauthorblockA{\textit{LITIS UR 4108}\\
\textit{ F-76000 Rouen, France} \\
}
}

\maketitle

\begin{abstract}
The emergence of 6G wireless networks promises to revolutionize vehicular communications by enabling ultra-reliable, low-latency, and high-capacity data exchange. In this context, collaborative perception techniques, where multiple vehicles or infrastructure nodes cooperate to jointly receive and decode transmitted signals, aim to enhance reliability and spectral efficiency for Connected Autonomous Vehicle (CAV) applications. In this paper, we propose an end-to-end wireless neural receiver based on a Differential Transformer architecture, tailored for 6G V2X communication with a specific focus on enabling collaborative perception among connected autonomous vehicles. Our model integrates key components of the 6G physical layer, designed to boost performance in dynamic and challenging autonomous driving environments. We validate the proposed system across a range of scenarios, including 3GPP-defined Urban Macro (UMa) channel. To assess the model’s real-world applicability, we evaluate its robustness within a V2X framework. In a collaborative perception scenario, our system processes heterogeneous LiDAR and camera data from four connected vehicles in dynamic cooperative vehicular networks. The results show significant improvements over state-of-the-art methods, achieving an average precision of 0.84, highlighting the potential of our proposed approach to enable robust, intelligent, and adaptive wireless cooperation for next-generation connected autonomous vehicles.
\end{abstract}


\section{Introduction}

Next-generation (6G) communication is expected to address connectivity issues, enabling seamless connections between connected devices \cite{1}. To achieve this, mmWave technology is anticipated to be utilized, providing increased data rates and millisecond (ms) latency. However, this introduces challenges in reconstruction of signal at the receiver due to the narrow beam of mmWave signals \cite{2}. To tackle this, Artificial Intelligence has been extensively explored in the literature for reconstructing received signals \cite{ribouh2024seecad}. Machine learning (ML) has gained prominence in wireless communication due to its ability to learn from data patterns and make autonomous decisions for unseen data \cite{3}.

As 6G technology advances, new challenges are emerging in mmWave signal reconstruction and multimodal data fusion, particularly in collaborative perception scenarios. For instance, in the Internet of Vehicles (IoV), devices need to efficiently exchange and process diverse environmental data to support real-time decisions \cite{a31}. This requires intelligent wireless networks capable of managing resources effectively while maintaining low latency and high  reliability \cite{a41} \cite{mvxvit}.

Recent advancements in ML have significantly enhanced communication systems. The use of ML to estimate conditionally Gaussian random vectors with structured covariance matrices for the channel has been explored in \cite{4}. Moreover, in \cite{5} the authors introduced an approximate message passing network with learned denoising approach to improve channel estimation in massive MIMO mmWave systems. Furthermore, in\cite{6} convolutional neural networks (CNNs) have been explored for equalization, that achieved significant reduction in error vector magnitude over traditional methods. In the context of demapping, the approach presented in \cite{7} leverage a deep neural network framework to efficiently compute bit Log-Likelihood Ratios (LLRs) for equalized symbols. Studies in \cite{8,9,10} have integrated deep learning into standard receiver processing, achieving notable performance improvements compared to the baseline approaches. Although the research works presented above in \cite{4,5,6,7,8,9,10} shows improved performance, yet their optimization remains limited to a single receiver operation.

In order to jointly optimize the receiver architecture, researchers in \cite{11} combined channel estimation and signal detection using a fully connected neural network that outperforms traditional MMSE based receivers. Furthermore, the method introduced in \cite{12} utilized CNNs to derive bit estimates directly from signal in the time-domain, showing superior performance compared to baseline in low to medium SNR conditions. In addition, a fully convolutional deep neural network method for 5G signal processing has been proposed in \cite{13}, achieving excellent performance in channel estimation and soft bit generation. In \cite{14} the authors highlights the challenge of domain generalization, where a neural receiver model trained on one channel model fails to perform well when tested on a different channel model. 

Differential Transformers, proposed in \cite{a14}, have emerged as a significant advancement in transformer models, operating on differential attention as its core learning algorithm. Their success lies in the ability of differential attention to reduce noise and focus on relevant context. One primary advantage of this approach is that it requires comparatively less data for training than conventional transformers \cite{a17}. This makes it suitable for AI-based receiver architectures \cite{a15} \cite{a16}

To this end, we propose a differential transformer-based neural receiver model for collaborative perception. The main contributions of this work are as follows:
\begin{itemize}
    \item We develop a novel neural receiver based on a Differential Transformer architecture for 6G V2X communication. The proposed receiver design is evaluated in an end-to-end wireless communication system across multiple scenarios, including 3GPP-defined Urban Macrocell (UMa) channel.
    \item We assess the robustness of the proposed receiver in a V2X framework by evaluating its performance in a collaborative perception scenario, where it processes heterogeneous data from four agents to emulate real-world V2X communication environment.

\end{itemize}
    
The remainder of this paper is organized as follows. Section~\ref{s3} presents the system model and the proposed solution. Section~\ref{s5} describes the experimental setup and scenarios. Section~\ref{s6} discusses the results, followed by Section~\ref{s7} that concludes the paper and outlines future research directions.

\section{System Model} \label{s3}
In this section, we present the system model for collaborative vehicular communication scenario involving multiple autonomous connected vehicles. These intelligent vehicles collaborate to enhance environmental perception through 6G-enabled V2X communication and end-to-end neural receivers based on a Differential Transformer architecture. As illustrated in Figure \ref{fig2}, at the transmitter side the system consists of two AVs (Vehicle 1 and Vehicle 2), Each vehicle independently captures its surroundings using onboard sensors (eg LIDAR and camera), generating raw perception data (detected nearby vehicles). Each AV, equipped with a communication pipeline comprising a channel encoder, a QAM modulator and an OFDM resource grid mapper, as described bellows:

\subsection{\textbf{At the transmitters (Vehicle 1 and Vehicle 2) }} \label{s31}

\subsubsection{Channel Encoder}

The captured data are converted into a binary bitstream then processed through a Low-Density Parity-Check (LDPC) encoder to add redundancy and improve robustness against transmission errors. The encoded data are expressed as follows:

\begin{equation}
    c = G \cdot b
\end{equation}

where, \( b \), \( G \) and \( c \) denote the information bits, the generator matrix and the encoded codewords, respectively.

\subsubsection{Mapping}
Subsequently, the encoded bits are mapped into complex 
symbols using a QAM modulator. the mapping process can be expressed as:

\begin{equation}
    x_i = \text{QAM}(c_i) = a_i + j b_i
\end{equation}

where, \( x_i \) is the QAM symbol corresponding to \( c_i \) and \( a_i, b_i \) are the real and imaginary components.

\subsubsection{Resource Grid Mapping}

The mapped QAM symbols are assigned to an OFDM resource grid as follows:

\begin{equation}
    X(m,n) = x_{m,n}
\end{equation}

\( X(m,n) \) refer to the OFDM resource element at subcarrier \( m \) and OFDM symbol \( n \), and \( x_{m,n} \) is the mapped QAM symbol.

The final time-domain OFDM signal is generated using the Inverse Fast Fourier Transform (IFFT):

\begin{equation}
    x(t) = \sum_{m=0}^{N-1} X(m,n) e^{j 2\pi m t / \zeta}
\end{equation}

where \( N \) is the Fast Fourier Transform size,\( \zeta \) represent the OFDM symbol duration and \( x(t) \) is the time-domain signal ready for transmission. The encoded and modulated data from each vehicle are transmitted through the wireless channel toward the receiver.
At the receiver side, an attention mechanism-based architecture is employed, incorporating a differential Transformer Block to process and extract semantic features from the received signal. The output is then passed through a channel decoder and a collaboration module to enhance the joint perception or decision-making between vehicles as decried bellows :
\subsection{\textbf{At the receiver (Vehicle 1, Vehicle 2, or Infrastructure)}} \label{s32}
The receiver is either one of the vehicles (Vehicle 1 or Vehicle 2) or a nearby edge server, such as a base station (BS). The receiver leverages a Diff-Transformer neural network model to perform channel estimation, equalization, and demapping of the received wireless signals.
\begin{figure*}[t]
  \centering
  \includegraphics[width=1\textwidth,height=0.25\textheight]{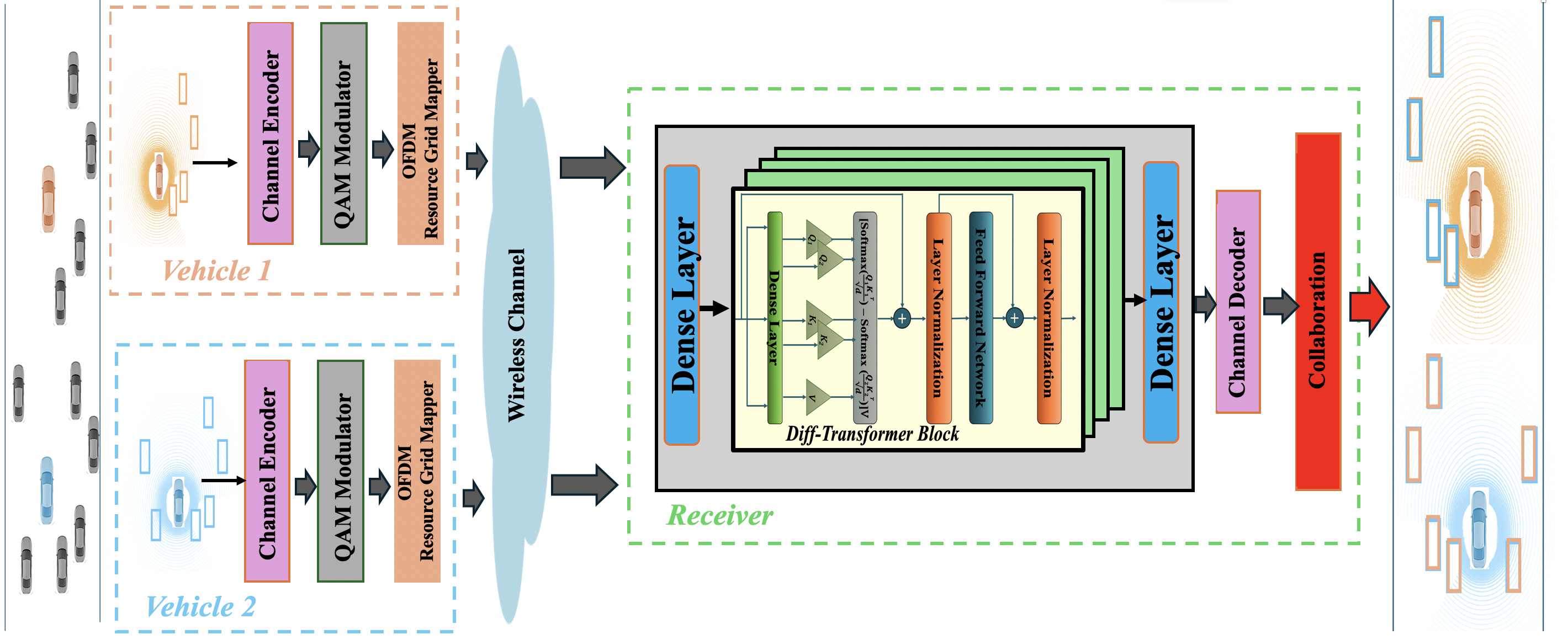}
  \caption{End-to-end collaborative perception framework based on Diff-Transformer-enabled wireless receiver }
  \label{fig2}
\end{figure*}
\subsubsection{Differential Transformer-Based Neural Receiver}
We propose a neural network designed to perform fundamental receiver functions aimed at reconstructing transmitted data. It processes the received signal as input and generates optimized log-likelihood ratios (LLRs) at the output. Our proposed architecture is built upon differential transformer encoder blocks\cite{a14}.  The Diff-Transformer block is a customized attention mechanism designed to enhance robustness to noise in communication signals. Unlike standard self-attention, it introduces a differential operation using two distinct sets of query-key projections ($Q_1, Q_2, K_1 $ and $K_2$), as shown in figure \ref{fig2}). This dual attention mechanism reduces sensitivity to noisy components by subtracting soft attention maps, thereby emphasizing consistent signal patterns. The differential attention output is defined as:

\begin{equation}
    \text{DiffAttn} = \left( \text{softmax} \left(\frac{Q_1 K_1^T}{\sqrt{d}}\right) - \text{softmax} \left(\frac{Q_2 K_2^T}{\sqrt{d}}\right) \right) V,    
\end{equation}
Where \( Q_1, Q_2, K_1, K_2 \in \mathbb{R}^{N \times d} \) and \( V \in \mathbb{R}^{N \times 2d} \). 

The proposed neural receiver architecture consists of an initial dense layer, followed by $4$ Differential Transformer blocks, and a final output dense layer. Each Differential Transformer block incorporates a differential attention mechanism with $4$ attention heads, $2$ normalization layers, and a feedforward sub-network with $2$  fully connected layers of dimension $128$.

The log-likelihood ratios (LLRs) are computed by processing the received signal \( Y_i \in \mathbb{R}^n \) through our proposed Diff-Transformer-based architecture, as described below. First, the signal is projected into a latent space using a fully connected layer:

\begin{equation}
    H^{(0)} = \phi(W_0 Y_i + b_0),
\end{equation}

where \( W_0 \in \mathbb{R}^{d_{\text{model}} \times n} \) and \( b_0 \in \mathbb{R}^{d_{\text{model}}} \) are the learnable weight matrix and bias vector, respectively, and \( \phi(\cdot) \) is the non-linear activation function. The resulting latent representation \( H^{(0)} \in \mathbb{R}^{N \times d_{\text{model}}} \) is passed through a stack of \( L \) Diff-Transformer blocks (\( L \) = 4).

Each Diff-Transformer block applies a differential attention mechanism, where the output of the \( l \)-th block is computed as:

\begin{equation}
    H^{(l)} = \phi(W_l \cdot \text{DiffAttn}(H^{(l-1)}) + b_l),
\end{equation}

where \( W_l \in \mathbb{R}^{d_{\text{model}} \times d_{\text{attn}}} \), \( b_l \in \mathbb{R}^{d_{\text{model}}} \), and \( \text{DiffAttn}(\cdot) \) denotes the differential attention operator applied to the previous hidden state \( H^{(l-1)} \). After \( L \) such transformations, the final latent representation \( H^{(L)} \in \mathbb{R}^{N \times d_h} \) is obtained.

Finally, this representation is linearly mapped to soft bit-level outputs:

\begin{equation}
    \hat{\text{LLR}} = W_{\text{LLR}} H^{(L)} + b_{\text{LLR}},
\end{equation}

where \( W_{\text{LLR}} \in \mathbb{R}^{N \times d_h} \) and \( b_{\text{LLR}} \in \mathbb{R}^{N} \) are trainable parameters. This produces a vector \( \hat{\text{LLR}} \in \mathbb{R}^{N} \), which contains the estimated log-likelihood ratios corresponding to the \( N \) transmitted bits.

\subsubsection{Decoding}

LDPC codes are decoded using an iterative belief propagation (BP). At each iteration \( t \), the LLR associated with a codeword bit \( c_i \) is updated as:

\begin{equation}
    L^{(t+1)}(c_i) = \hat{\text{LLR}}(c_i) + \sum_{j \in \mathcal{N}(i)} \beta_{j \to i}^{(t)},
\end{equation}

where \( \hat{\text{LLR}}(c_i) \in \mathbb{R} \) denotes the estimated log-likelihood ratio produced by the neural receiver , and \( \beta_{j \to i}^{(t)} \) represents the extrinsic message passed from the \( j \)-th check node to the \( i \)-th variable node in the Tanner graph at iteration \( t \). The set \( \mathcal{N}(i) \) refers to the indices of all check nodes connected to variable node \( i \).

After a fixed number of BP iterations or upon convergence, a hard decision is applied to each bit as:

\begin{equation}
    \hat{b}_i =
    \begin{cases} 
    1, & L^{(T)}(c_i) < 0 \\
    0, & L^{(T)}(c_i) \geq 0
    \end{cases}
\end{equation}

where \( L^{(T)}(c_i) \) is the final updated LLR after \( T \) iterations, and \( \hat{b}_i \in \{0,1\} \) is the decoded bit.

\subsection{Collaborative Perception}

Once the channel decoder recovers the perception data sent by Vehicle 1 and Vehicle 2, the system performs collaborative fusion to build a unified and accurate environmental representation. This fusion enhances situational awareness and supports better decision-making. 

We adopt the HEAL (HEterogeneous ALliance) framework \cite{HEAL} , which transforms the reconstructed sensor data (LiDAR and images) into a bird’s-eye view (BEV) format for spatial alignment. A Pyramid Fusion module then aggregates BEV features across multiple scales, prioritizing semantically important regions, such as areas directly in front of the vehicles.

The aligned features are merged into a single global map, improving spatial coverage, reducing occlusions, and enhancing object detection. The final fused perception output is then used by the receiving vehicle or distributed by the edge server to all involved agents to support safer and more efficient autonomous navigation.

\section{Experimental Setup} \label{s5}
\vspace{-0,1em}
To evaluate the proposed system, we design two complementary experiments levels : one at the physical layer to assess communication performance using our neural receiver, and another at the perception level to evaluate the benefits of collaborative multi-agent fusion in a V2X environment.
\subsection{Physical Layer-Based Evaluation}
We assess the performance of the proposed neural receiver through an end-to-end OFDM-based 6G communication system. The setup includes 128 subcarriers with a subcarrier spacing of 240 kHz, and 14 OFDM symbols per frame arranged using a Kronecker pilot pattern. Each symbol carries 6 bits encoded via 64-QAM modulation. This setup was implemented using the Sionna framework \cite{15}
The sensor-captured data are transmitted from vehicles and propagates through a wireless Urban Macrocell (UMa) channel, as defined by 3GPP standards \cite{17}, capturing realistic conditions including path loss and shadow fading. At the receiver side, as described in Section \ref{s3}, the received signals are processed using the pre-trained Diff-Transformer-based neural receiver to estimate log-likelihood ratios (LLRs), which are then passed through an LDPC decoder to recover the original bitstream. The Bit Error Rate (BER) is computed by comparing the decoded bits against the transmitted ground truth.
\\The proposed Diff-Transformer-based neural receiver model was trained on 12 million  data samples. During training, different levels of SNR and channel topologies based on the UMa model were used to expose the network to a diverse range of conditions. Binary cross-entropy (BCE) loss was used as the optimization function, and performance was monitored using accuracy and loss metrics to ensure convergence and generalization.

\subsection{Collaborative Perception-Based Evaluation}
To validate the effectiveness of collaborative perception in a V2X context, we employ the HEAL framework \cite{HEAL}, which enables fusion of heterogeneous sensor data across multiple agents. Four types of perception agents are considered: two LiDAR-based models : $L_P^{(64)}$ (64-channel, PointPillars), $L_S^{(32)}$ (32-channel,SECOND) and two camera-based models : $C_E^{(384)}$ (384-pixel height, EfficientNet encode), $C_R^{(336)}$ (336-pixel height, ResNet encode).

Following the protocol described in \cite{HEAL}, $L_P^{(64)}$ serves as the collaboration base, and its features are fused using the Pyramid Fusion module. The additional agents are incrementally integrated into this shared feature space. The collaboration base is trained on the OPV2V dataset, while the new agents are trained using the OPV2V-H dataset. Each model is trained using the Adam optimizer.

Training is conducted on an NVIDIA Tesla P40 GPU with 24 GB of memory. The receiver (vehicle or edge server) hosts the collaborative model and performs feature fusion using data transmitted by the connected vehicles. Prior to fusion, the received image and LiDAR data are reconstructed using the proposed Diff-Transformer Neural Receiver.
\begin{figure}[b]
  \centering
  \includegraphics[width=0.50\textwidth,height=0.25\textheight]{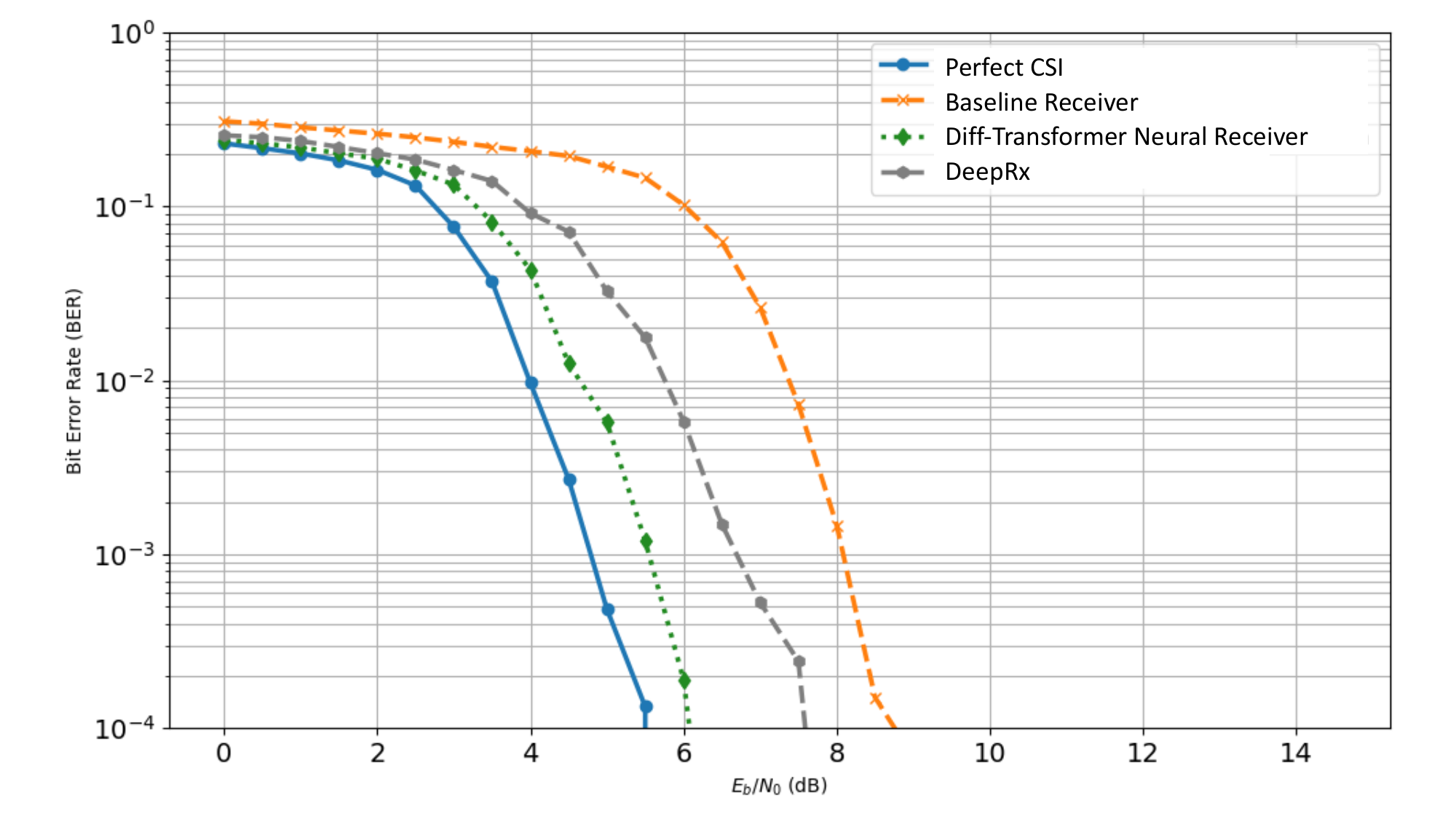}
  \caption{BER vs. SNR performance over a UMa channel with vehicle speeds ranging from 0 to 60 km/h.}
  \label{fig4}
\end{figure}
\section{Results and Discussion} \label{s6}
This section presents the comprehensive evaluation results  obtained from the two complementary evaluation levels. At the physical layer, we assess the communication performance of the proposed Diff-Transformer-based neural receiver. At the perception layer, we evaluate the impact of enhanced signal recovery on downstream multi-agent perception tasks, focusing on object detection accuracy and collaborative fusion performance. The performance of the proposed Diff-Transformer-based neural receiver is compared with the following state-of-the-art techniques: \textbf{(i)} \textit{Perfect CSI}, which assumes ideal and error-free CSI; \textbf{(ii)} \textit{Baseline}, conventional receiver  and \textbf{(iii)} \textit{DeepRx} proposed in \cite{13}, re-trained on the same dataset and under identical channel conditions for a fair comparison with the proposed method.

\subsection{At the Physical layer}
\begin{figure}[t]
  \centering
  \includegraphics[width=0.5\textwidth,height=0.25\textheight]{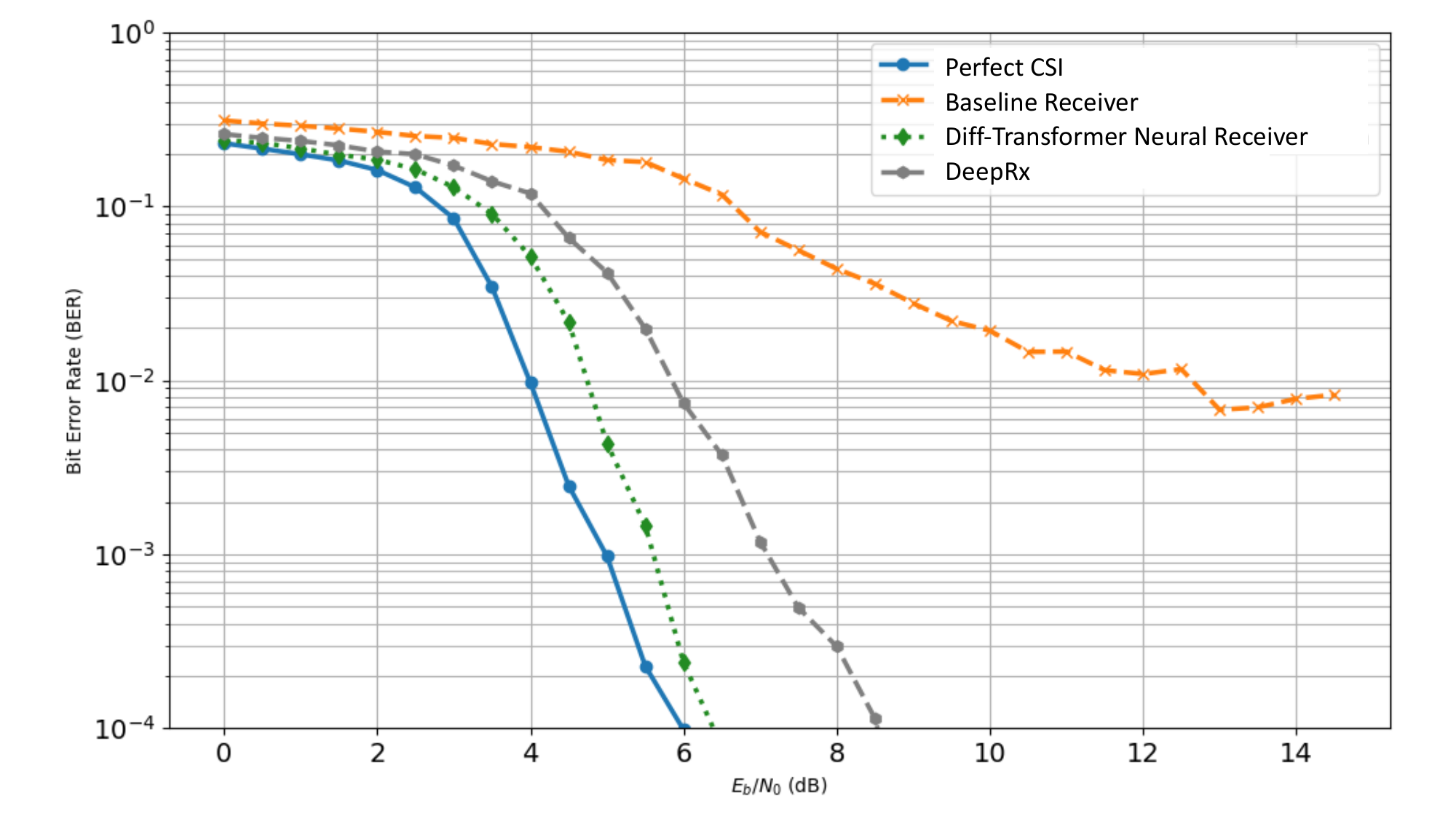}
  \caption{BER vs. SNR performance over a UMa channel with vehicle speeds ranging from 60 to 120 km/h. }
  \label{fig3}
\end{figure}

Bit error rate (BER) was computed across multiple signal-to-noise ratio (SNR) values to evaluate the performance of each receiver model under realistic communication conditions. The transmission scenario assumes that the vehicles operate in an urban environment, moving at speeds ranging from 0 to 60 km/h. As shown in Fig.~\ref{fig4}, the proposed Diff-Transformerr-based neural receiver outperforms both the CNN-based and baseline models. At an SNR of 6 dB, the proposed model achieves a BER of approximately \(2 \times 10^{-4}\), while the CNN-based receiver and the baseline yield BERs of \(1.2 \times 10^{-3}\) and \(1 \times 10^{-1}\), respectively. These values correspond to relative BLER reductions of approximately 83\% compared to the CNN-based model and 99.8\% relative to the baseline. Additionally, to reach a target BER of \(10^{-4}\), the proposed model requires only 5.75~dB, whereas the CNN-based and baseline receivers require 7.5 dB and over 10 dB, respectively. These significant improvements confirm the effectiveness of differential attention in enhancing communication robustness under challenging urban conditions.
To further evaluate generalization across mobility profiles, we consider an urban transmission scenario with high vehicle speeds ranging from 60 to 120 km/h.  
As shown in Fig.~\ref{fig3}, the proposed Diff-transformer-based neural receiver, continues to outperform both the baseline receiver and the DeepRx. The BER of the proposed model converges to a value close to zero at an SNR of 6.25 dB, indicating highly reliable bit recovery under degraded channel conditions. In contrast, the CNN-based model requires an SNR of 8.25 dB to reach a comparable BER, while the baseline model converge to similar performance at approximately 20 dB. These results demonstrate performance gains of 2.25 dB and 13.75 dB over the CNN-based receiver  and baseline receiver, respectively. These results reinforce the robustness of the proposed model under high-mobility conditions.

\begin{figure}[t!]
  \centering
  \includegraphics[width=0.5\textwidth,height=0.25\textheight]{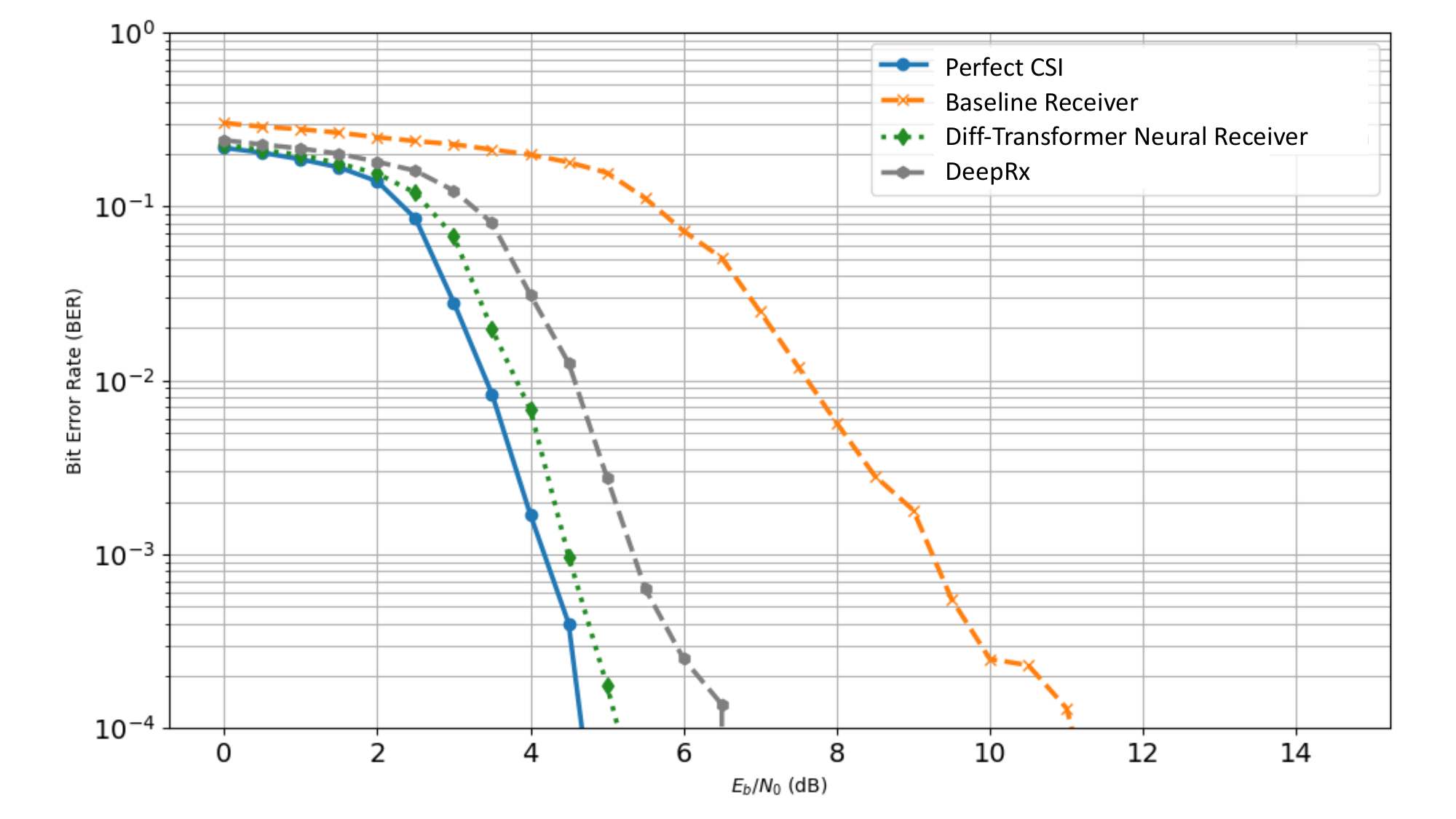}
  \caption{BER vs. SNR performance over a CDL channel with vehicle speeds ranging from 60 to 120 km/h.}
  \label{fig5}
\end{figure}
Additionally, to assess the  capability of the proposed neural receiver under other wireless channel conditions, we evaluated its performance on delay line (CDL) channel to emulate dynamic channel variations caused by high-mobility. As illustrated in Fig.~\ref{fig5}, the proposed Diff transformer-based neural receiver consistently outperforms both the DeepRx and the baseline approach across varying SNR values. The proposed model demonstrates superior performance, achieving minimal BER with a 0.5~dB SNR advantage over the DeepRx and a 5~dB improvement compared to the baseline. These results highlight the robustness of the proposed architecture under unseen, time and frequency  selective fading conditions typical of real-world vehicular environments.
\subsection{At the perception layer}
\begin{figure}[h]
    \centering
    \begin{subfigure}[t]{1\linewidth}
        \centering
        \includegraphics[width=\linewidth,height=0.5\linewidth]{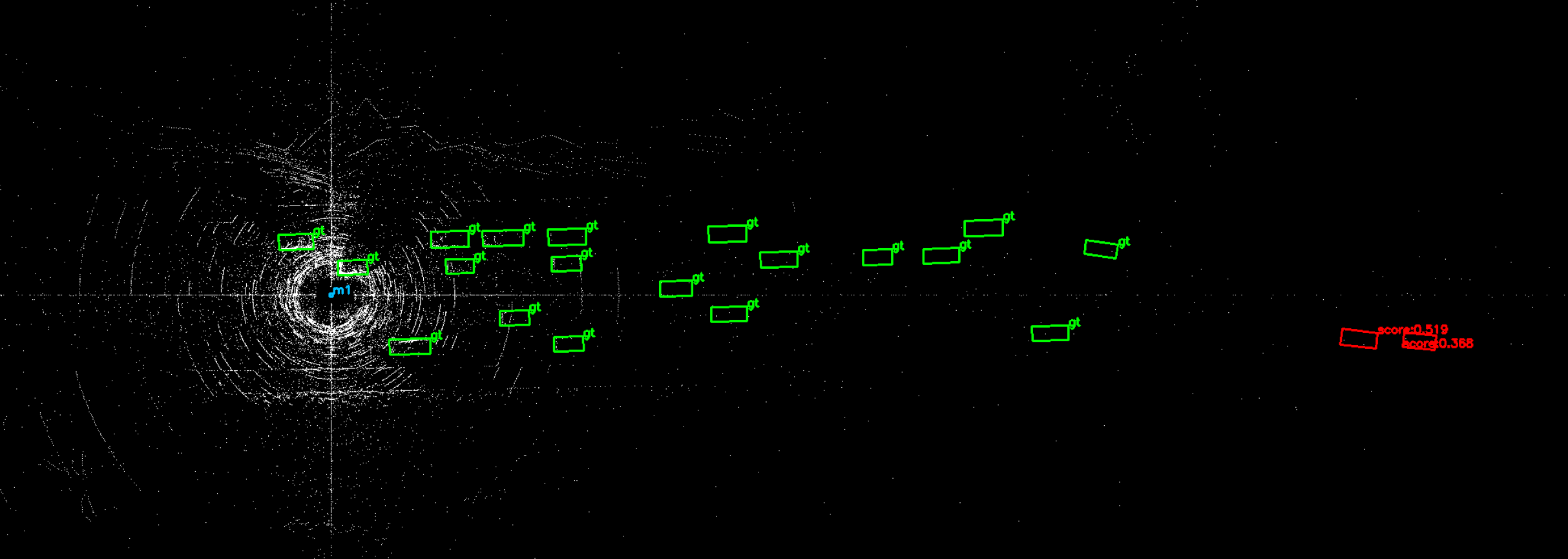}
        \caption{Baseline Receiver}
        \label{fig:obj_det_baseline}
    \end{subfigure}
    \hfill
    \begin{subfigure}[t]{1\linewidth}
        \centering
        \includegraphics[width=\linewidth,height=0.51\linewidth]{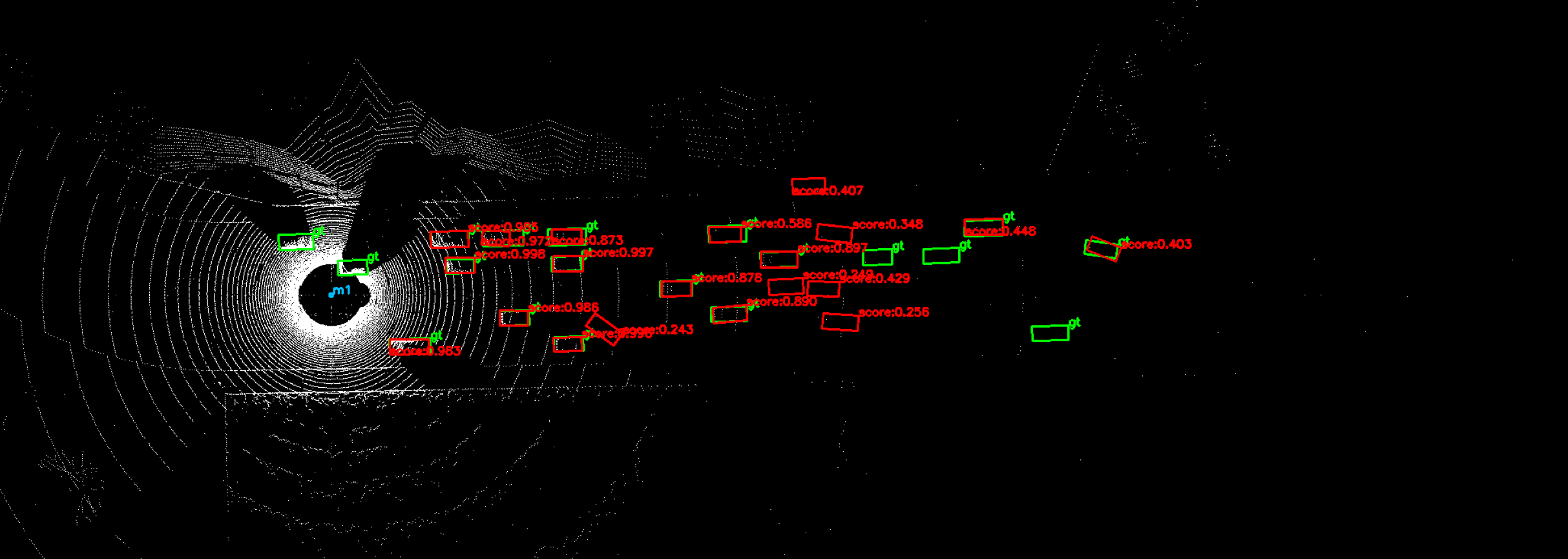}
        \caption{DeepRx}
        \label{fig:obj_det_deeprx}
    \end{subfigure}
    \hfill
    \begin{subfigure}[t]{1\linewidth}
        \centering
        \includegraphics[width=\linewidth,height=0.51\linewidth]{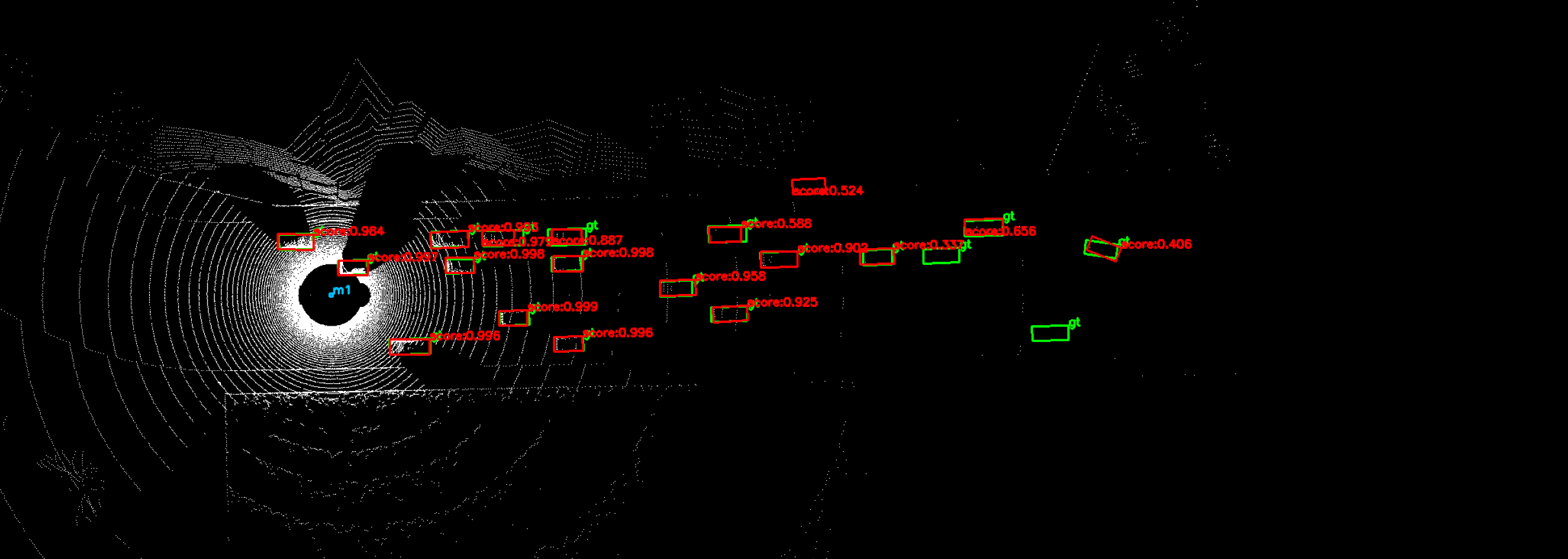}
        \caption{Diff-Transformer}
        \label{fig:obj_det_difftrans}
    \end{subfigure}
    \caption{Visualization Results for Object Detection with 4 Heterogeneous Agents. Sub-figures show \textcolor{red}{predicted} and \textcolor{green}{ground truth} bounding boxes for (a) Baseline, (b) DeepRx, and (c) Diff-Transformer Neural Receiver.}
    \label{fig:obj_det_combined}
\end{figure}

To investigate the robustness of our Diff-Transformer Neural Receiver within V2X collaborative perception framework, we evaluate its efficacy in reconstructing heterogeneous perception data from four agents, as detailed in Section~\ref{s5}. Object detection performance is quantified using Average Precision (AP) at Intersection over Union (IoU) thresholds of 30, 50, and 70 denoted as AP30, AP50 and AP70, respectively. Table~\ref{tab:ap_results} summarizes the object detection performance of the proposed Diff-Transformer Neural Receiver in comparison with the Baseline and DeepRx benchmarks. The proposed model achieves substantially higher Average Precision (AP) scores 0.89, 0.89, and 0.84 for AP30, AP50, and AP70, respectively outperforming DeepRx (0.73, 0.73, 0.61) and the Baseline (0.00, 0.00, 0.00). This performance margin, particularly evident at the AP70 threshold, underscores the robustness of the proposed model in reconstructing complex, heterogeneous perception data from LiDAR and camera inputs within the HEAL framework. Visualization of these results, depicted in Fig.~\ref{fig:obj_det_combined}, showcases \textcolor{red}{predicted} and \textcolor{green}{ground truth} bounding boxes for (a) Baseline, (b) DeepRx, and (c) Diff. Transformer. The sub-figures reveal that Diff. Transformer’s predicted boxes closely align with ground truth, unlike Baseline's complete failure and DeepRx's moderate inaccuracies, affirming its suitability for real-world V2X applications.
\renewcommand{\arraystretch}{1.5}
\begin{table}[t]
\centering
\caption{Average Precision (AP) at IoU Thresholds for Collaborative Perception with Four Heterogeneous Agents}

\label{tab:ap_results}
\begin{tabular}{lccc}
\hline

\textbf{Receiver} & \textbf{AP30} & \textbf{AP50} & \textbf{AP70} \\
\hline
Baseline Receiver & 0.00 & 0.00 & 0.00 \\
DeepRx~\cite{13} & 0.73 & 0.73 & 0.61 \\
Diff-Transformer (Proposed) & \textbf{0.89} & \textbf{0.89} & \textbf{0.84} \\
\hline
\end{tabular}
\end{table}

\section{Conclusion and Future Work} \label{s7}
This paper presented a Differential Transformer-based neural receiver designed to enhance communications performance in high-mobility scenarios and multi-agent perception systems. The proposed architecture integrates a differential attention mechanism to improve robustness against channel variability and noise, and was evaluated across physical and perception layers. Experimental results demonstrated that the model outperforms conventional and CNN-based receivers in terms of BER over UMa Channel, achieving significant gains in overall system performance. Furthermore, in In the context of V2X collaborative perception, the proposed receiver demonstrates robust capabilities in reconstructing heterogeneous sensor data from multiple agents, resulting in high object detection performance across all Intersection over Union (IoU) thresholds. These results underscore the effectiveness of the model in enabling intelligent, reliable communication systems for future 6G vehicular network. Future research will focus on integrating the receiver with real-time vehicular platforms and deploying it on hardware to assess its performance under  real-world conditions.

\end{document}